\documentclass[12pt,a4paper]{article}%
\usepackage{graphicx}
\usepackage{amsmath}
\usepackage{amsfonts}
\usepackage{amssymb}%
\begin{document}

\title{Dynamically generated embeddings of spacetime}
\author{F. Dahia$^{1}$ and C. Romero$^{2}$\\$^{1}$Departamento de F\'{\i}sica, UFCG,\\Campina Grande-PB, Brazil, 58109-970.\\$^{2}$Departamento de F\'{\i}sica, UFPB,\\Cx. 5008, 58051-970, Jo\~{a}o Pessoa -PB, Brazil}
\maketitle

\begin{abstract}
We discuss how embeddings in connection with the Campbell-Magaard (CM) theorem
can have a physical interpretation. We show that any embedding whose local
existence is guaranteed by the CM theorem can be viewed as a result of the
dynamical evolution of initial data given in a four-dimensional spacelike
hypersurface. By using the CM theorem, we establish that for any analytic
spacetime, there exist appropriate initial data whose Cauchy development is a
five-dimensional vacuum space into which the spacetime is locally embedded. We
shall see also that the spacetime embedded is Cauchy stable with respect these
the initial data.

\end{abstract}

The development of the braneworld scenario \cite{add1,add2,rs1,rs2}, in which
our ordinary spacetime is viewed as a hypersurface of a higher-dimensional
space, has greatly contributed to increase recent interest on embeddings
theorems. The so-called non-compact Kaluza-Klein (NKK) models
\cite{overduim,ponce,book} have also motivated the study of embedding problems
of the spacetime.

In this context, the Campbell-Magaard (CM) theorem \cite{campbell,
magaard,romero} and its variants \cite{anderson,dahia1,dahia3,dahia2,seahra}
are of special interest when the embedding spaces possess only one extra
dimension as is the case of the Randall-Sundrum braneworld scenario
\cite{rs1,rs2} and the NKK models \cite{overduim,ponce,book}.

The CM theorem ensures the existence of local and analytic isometric
embedding\footnote{Henceforth embedding for us will mean analytic and
isometric embedding.} of any $n$-dimensional analytic manifold $M$ into a
Ricci-flat $\left(  n+1\right)  $-dimensional space. It can be stated as
follow: Given a $n$-dimensional analytic space $\left(  M,g\right)  $, where
$g$ stands for the metric, for any point \ $p\in M$ there exists a vacuum
solution of Einstein equations in $\left(  n+1\right)  $-dimensions into which
a neighborhood of $p$ in $M$ can be embedded. The CM theorem is valid for any
analytic metric $g$ irrespective of its signature and for dimension $n\geq2$.
However, having in mind applications to physics, we shall consider $n=4$,
throughout the paper, and assume that the metric is Lorentzian. The CM theorem
is local, but the extension to the case of global embedding has been studied
\cite{global}\emph{.}

Recently some interesting issues concerning the physical interpretation of
embeddings in the light of the CM theorem have been arisen
\cite{anderson1,reza}. It has been argued that, since the CM theorem is based
on the Cauchy-Kowalewski theorem it disregards causality and cannot guarantee
a continuous dependence between the metric of ambient space and spacetime
metric. Therefore, by lacking these important properties, the\emph{
}embeddings\emph{ }obtained by employing the CM theorem would have no physical meaning.

In this paper we intend to show that, as a matter of fact, ambient spaces
whose local existence is guaranteed by the CM theorem possess, in a certain
domain, those desirable physical properties with respect to appropriate
initial\emph{ }data. Indeed, we shall show that each five-dimensional vacuum
space into which the spacetime can be locally embedded may be locally viewed
as the Cauchy development of some initial data given on a spacelike
four-dimensional hypersurface and hence satisfies the stability and causality
conditions with respect to these initial data. Further, we shall see that the
region of $M$ which is embedded in the Cauchy development of an initial data
set possesses also Cauchy stability and domain of dependence property with
respect to these initial data.

In order to appropriately address these issues, let us begin our discussion by
considering a brief sketch of Magaard's proof of the CM\ theorem.

\section{The Campbell-Magaard theorem}

Consider the Lorentzian metric of the five-dimensional space written in a
Gaussian form
\begin{equation}
ds^{2}=\overline{g}_{ij}\left(  x,\psi\right)  dx^{i}dx^{j}+d\psi^{2},
\label{hds2}%
\end{equation}
where $x=\left(  x^{1},...,x^{4}\right)  $, and Latin indices run from $1$ to
$4$ while the Greek ones go from $1$ to $5.$

By splitting the vacuum Einstein equations in terms of the extrinsic and
intrinsic curvatures of the slices $\psi=const$, it can be shown that the
equations have the following structure:%

\begin{align}
\frac{\partial^{2}\overline{g}_{ij}}{\partial\psi^{2}} &  =F_{ij}\left(
\overline{g},\frac{\partial\overline{g}}{\partial\psi},\frac{\partial
\overline{g}}{\partial x},\frac{\partial^{2}\overline{g}}{\partial x^{2}%
},\frac{\partial^{2}\overline{g}}{\partial x\partial\psi}\right)
\label{dyn}\\
\nabla_{j}\left(  \Omega^{ij}-g^{ij}\Omega\right)   &  =0\label{c1}\\
R+\Omega^{2}-\Omega_{ij}\Omega^{ij} &  =0,\label{c2}%
\end{align}
where $F_{ij}$ are analytic functions of their arguments, $\nabla_{j}$ is the
covariant derivative with respect to the induced metric $g_{ij}=\overline
{g}_{ij}\left(  x,\psi=const\right)  $; $R$ and\ $\Omega_{ij}$\ denote,
respectively, the scalar curvature and the extrinsic curvature of the
hypersurface $\psi=const;$ and $\Omega=g^{ij}\Omega_{ij}$. Recall that in the
coordinates adopted the extrinsic curvature assumes the simple form:
\begin{equation}
\Omega_{ij}=-\frac{1}{2}\frac{\partial\overline{g}_{ij}}{\partial\psi
}.\label{ext}%
\end{equation}
It is well known that, owing to the Bianchi identities, the second and third
equations need to be imposed only on the hypersurface, since they are
propagated by the first one. In this sense, it is said that the Einstein
equations consist of the \textit{dynamical }equation (\ref{dyn}) plus
\textit{constraint equations }(\ref{c1}) and (\ref{c2}) for $\Omega_{ij}$ and
$g_{ij}$ .

Let now consider the hypersurface $\psi=0.$ According to the Cauchy-Kowalewski
theorem, for any point in this hypersurface, say the origin, there is an open
set in five dimensions containing that point, where the equation (\ref{dyn})
always has a unique analytic solution $\overline{g}_{ik}\left(  x,\psi\right)
$ provided that the following analytic initial conditions are specified:
\begin{align}
\overline{g}_{ij}\left(  x,0\right)   &  =g_{ij}\left(  x\right)
\label{cig}\\
\left.  \frac{\partial\overline{g}_{ij}}{\partial\psi}\right\vert _{\psi=0} &
=-2\Omega_{ij}\left(  x\right)  .\label{cih}%
\end{align}

From the perspective of the embedding problem these initial conditions
represent, respectively, the metric and the extrinsic curvature of the
hypersurface $\psi=0$, whereas the solution of equation (\ref{dyn}) gives the
metric of the $\left(  n+1\right)  -$dimensional space. Thus, if there is a
solution for the constraint equations for any given metric $g_{ik}$, then the
theorem is proved, since the solution found $\overline{g}_{ij}\left(
x,\psi\right)  $ substituted in (\ref{hds2}) will give rise to a metric that
satisfies the vacuum Einstein equation $R_{\mu\nu}=0$. Clearly, the embedding
map is then given by the equation $\left(  x,\psi=0\right)  $.

It turns out, as Magaard has proved \cite{magaard,dahia1}, that the constraint
equations always have a solution. Indeed, by simple counting operation we can
see that there are $n(n+1)/2$ unknown functions (the independent elements of
extrinsic curvature) and $n+1$ constraint equations. The metric $g_{ij}\left(
x\right)  $\ must be considered as a given datum. For $n\geq2$, the number of
variables is equal or greater than the number of equations. Magaard has shown
that after the elimination of equation (\ref{c2}), the first-order
differential equation (\ref{c1}) can be written in a canonical form (similar
to (\ref{dyn})) with respect to $n$ components of $\Omega_{ij}$ conveniently
chosen. Taking initial conditions for these components in such a way that the
right-hand side of the mentioned equation is analytic at the origin, the
Cauchy-Kowalewski theorem can be applied once more to guarantee the existence
of a solution for the constraint equations.

\section{Dynamically generated embedding}

From the above we see that Magaard's proof of the CM theorem is formulated in
terms of an initial value problem. Roughly speaking we can say that a
$(3+1)$-spacetime is taken as part of the initial data and it is
\textquotedblleft propagated\textquotedblright\ along a spacelike extra
dimension by the dynamical part of the Einstein vacuum equations to generate
the higher-dimensional space. Nevertheless, it is enough clear that, despite
some similarities, the CM theorem is not concerned with real dynamical
propagation since the initial data \textquotedblleft evolve\textquotedblright%
\ along a spacelike direction. Therefore there is no reason why we should
expect a causality relation between different slices of the higher dimensional space.

However, we can look at this picture from a different perspective. Indeed,
supported by the CM theorem, we know that given any point $p\in$ $M$ there is
a five-dimensional vacuum space $(\widetilde{M},\widetilde{g})$ into which a
neighborhood of $p$ in $M$ is embedded. Now we can determine an open subset of
$\widetilde{M}$, say, $\widetilde{O}$, containing the point $p$ in which there
exists a four-dimensional hypersurface $\Sigma$, which is spacelike
everywhere, acausal and that contains the point $p$ (see appendix I). The
embedding of $\Sigma$ into $\left(  \widetilde{O},\widetilde{g}\right)  $
induces a positive definite metric $h$ in the hypersurface. Let $K$ be the
extrinsic curvature of $\Sigma$ in $\left(  \widetilde{O},\widetilde
{g}\right)  .$ The metric and the extrinsic curvature are analytic fields in
the hypersurface $\Sigma$, thus they belong to the local Sobolev
space\footnote{By writing $h\in W^{m}\left(  \Sigma\right)  $ we mean that the
norm of $h$ together with its covariant derivatives of order equal or less
than $m$ are square integrable in any open set $\mathcal{U}$ of $\Sigma$ with
compact closure. For the sake of simplicity, we shall assume that the norm and
derivatives are calculated with respect to an Euclidean metric. Here we are
adopting the notation of Sobolev spaces used in \cite{HE}.}\emph{ }%
$W^{m}\left(  \Sigma\right)  $ for any $m$. The set $\left(  h,K,\Sigma
\right)  $ constitutes appropriate initial data for the Einstein vacuum
equations, since $h$ and $K$ satisfy the vacuum constraint equations in the
hypersurface $\Sigma$ and fulfill the required condition of regularity (see
\cite{HE}, page 248-249, and \cite{CB}, for instance).

Consider now $D\left(  \Sigma\right)  $, the domain of dependence of $\Sigma$
relative to $\left(  \widetilde{O},\widetilde{g}\right)  $ \cite{wald}. Since
$\Sigma$ is an acausal hypersurface of $\widetilde{O}$, then $D\left(
\Sigma\right)  $ is open in $\widetilde{O}$ (see \cite{oneill}, page 425). Of
course $D(\Sigma)$ is a non-empty set, since $\Sigma\in D(\Sigma).$ By
construction, the five-dimensional manifold $(D\left(  \Sigma\right)
,\widetilde{g})$ is a solution of the Einstein vacuum equations, hence
$(D\left(  \Sigma\right)  ,\widetilde{g})$ is a Cauchy development for the
Einstein vacuum equations of the initial data $\left(  h,K,\Sigma\right)  .$

As we have mentioned before $D(\Sigma)$ is open, and thus the non-empty set
$M\cap D\left(  \Sigma\right)  $ is a neighbourhood of $p$ in $M$ contained in
$D\left(  \Sigma\right)  $. Therefore $M\cap D\left(  \Sigma\right)  $ is
embedded in $D\left(  \Sigma\right)  $, i.e., in a Cauchy development of
$\left(  h,K,\Sigma\right)  $ (see figure 1). In other words the dynamical
evolution of the initial data $\left(  h,K,\Sigma\right)  $ generates a
five-dimensional \ vacuum space into which the spacetime is locally embedded.
In this sense, we can say that this local embedding is dynamically generated
by the physical propagation of those initial data.

More precisely this result can be stated as follows: \textit{Consider an
analytic spacetime }$\left(  M,g\right)  .$\textit{\ For any }$p\in
M$\textit{\ there are initial data }$\left(  h,K,\Sigma\right)  $%
\textit{\ whose Cauchy development for the Einstein vacuum\ equations is a
five-dimensional vacuum space into which a neighbourhood of }$p$\textit{ in
}$M$\textit{\ is analytically and isometrically embedded.}

\begin{figure}[ptb]
\begin{center}
\includegraphics[scale=0.4]
{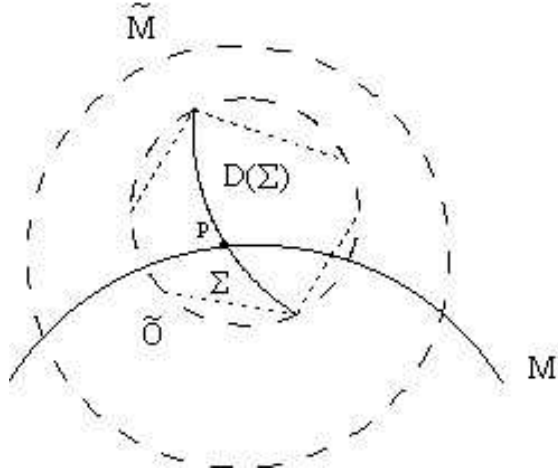}
\end{center}
\caption{{\small {Sketch of the local embedding of spacetime $M$ into the
Cauchy development $D\left(  \Sigma\right)  $ of initial data given in an
acausal spacelike four dimensional hypersurface $\Sigma.$}}}%
\end{figure}

Furthermore, the Einstein vacuum\ equations admit a well-posed initial value
formulation with respect to the data $\left(  h,K,\Sigma\right)  $ (see, for
example,\cite{HE,CB,wald,Friedrich}). Therefore the general properties of
solutions of the vacuum Einstein equations, related to the hyperbolic
character of the differential equations, are applicable to our solution
$\left(  D\left(  \Sigma\right)  ,\widetilde{g}\right)  $. This ensures that
the dependence of the solution $\left(  D\left(  \Sigma\right)  ,\widetilde
{g}\right)  $ on the initial data $\left(  h,K,\Sigma\right)  $ is continuous
(Cauchy stability). As a consequence the spacetime embedded into $\left(
D\left(  \Sigma\right)  ,\widetilde{g}\right)  $ is stable in a similar sense
too, as we describe in the next section.

Another important property is that any change of data outside $\Sigma$ does
not affect the solution in the future domain of dependence (causality). Thus
it follows that any perturbation outside $\Sigma$ will not disturb the
embedding of spacetime in $D\left(  \Sigma\right)  $.

\section{Cauchy Stability}

Consider an analytic spacetime $\left(  M,g\right)  $ and let $\left(
h,K,\Sigma\right)  $ be a set of analytic initial data with a Cauchy
development $\left(  D\left(  \Sigma\right)  ,\widetilde{g}\right)  $ in which
the spacetime is locally embedded, around $p\in M.$ Additionally let us admit
that this initial data set satisfies the following property: the image of $p$
through the embedding lies in $\Sigma$. In other words, corresponding to the
set $\left(  h,K,\Sigma\right)  $ there are some neighbourhood $O$ of $p$ in
$M$ and a map $\varphi:O\subset M\rightarrow$ $D\left(  \Sigma\right)  $ which
is an embedding, with $\varphi\left(  p\right)  \in\Sigma$ \footnote{The
existence of the set $\left(  h,K,\Sigma\right)  $ was shown in the previous
section. Possibly the initial data set is not unique. The results obtained in
this section are applicable separately to each one of all possible initial
data set.}.

Now we denote by $\left(  h^{\prime},K^{\prime}\right)  $ a new set of initial
data which satisfies the vacuum Einstein constraint equation in $\Sigma$. For
the sake of simplicity let us assume that the fields $h^{\prime}$ and
$K^{\prime}$ are $C^{\infty}$ in $\Sigma$. In this case, the new generated
metric $g^{\ast}$ is a $C^{\infty}$ field.

Let $V$ be an open set of $J^{+}\left(  \Sigma\right)  $, the causal future of
$\Sigma$ in $\left(  D\left(  \Sigma\right)  ,\widetilde{g}\right)  $, with
compact closure and $\mathcal{U}$ $\subset\Sigma$ some neighbourhood of
$J^{-}\left(  \overline{V}\right)  \cap\Sigma$, the causal past of
$\overline{V}$ (the closure of $V$) in $\Sigma$, with compact closure in
$\Sigma$. According to the Cauchy stability theorem (see \cite{HE}, page 253,
and \cite{CB}), for any $\varepsilon>0$ there is some $\delta>0$ such that any
initial data $\left(  h^{\prime},K^{\prime}\right)  $ on $\Sigma$ close to
$\left(  h,K\right)  $ in $\mathcal{U}$ with respect to the local Sobolev
norm, i.e., $\left\|  h^{\prime}-h,\mathcal{U}\right\|  _{m}^{\sim}<\delta$
and $\left\|  K^{\prime}-K,\mathcal{U}\right\|  _{m-1}^{\sim}<\delta$ ( $m>4)$
\footnote{By the symbol \symbol{126}we mean that derivatives are taken only in
tangent directions of $\Sigma$.}, give rise to a new metric $g^{\ast}$ which
is near to the old one $\widetilde{g}$ in $V$, i.e., $\left\|  g^{\ast
}-\widetilde{g},V\right\|  _{m}<\varepsilon$.

Now let $V$ be such that $V$ $\cap\left(  M\cap D\left(  \Sigma\right)
\right)  =N$ is a non-empty set, where $M\cap D\left(  \Sigma\right)  $ means
$\varphi\left(  O\right)  ,$ the image of $O$ through the embedding. And let
$g^{\prime}$ be the induced metric on $N$ by the embedding of $N$ in $\left(
V,g^{\ast}\right)  $. We shall see that if $\delta$ is sufficiently small then
$g^{\prime}$ will be a Lorentzian metric and it will be close to the spacetime
metric $g$ in $N$.

For the sake of simplicity, let us make some assumptions. First, we assume
that $D\left(  \Sigma\right)  $ is covered by Gaussian coordinates
(\ref{hds2}) adapted to the embedding, in which the embedding map is $\left(
x,\psi=0\right)  $. If this was not the case, we could find a neighbourhood
$S$ of $p$ in $\Sigma$ and a neighbourhood $\widetilde{O}$ of $p$ in $D\left(
\Sigma\right)  $ such that the domain of dependence of $S$ relative to
$\left(  \widetilde{O},\widetilde{g}\right)  $, $D\left(  S,\widetilde
{O}\right)  ,$ is covered by (\ref{hds2}). We would proceed in the following
manner. Since the embedding exists, we know that $M\cap D\left(
\Sigma\right)  $ is a timelike hypersurface of $\left(  D\left(
\Sigma\right)  ,\widetilde{g}\right)  .$ By the usual procedure we construct,
from the geodesics that cross $M\cap D\left(  \Sigma\right)  $ orthogonally,
Gaussian normal coordinates in a neighbourhood $\widetilde{O}$ of $p$ in
$D\left(  \Sigma\right)  .$ Now make $S=\Sigma\cap\widetilde{O}$. We then
concentrate our analysis in the region $D\left(  S,\widetilde{O}\right)  $.

Second, let us assume that the Sobolev norm is evaluated with respect to an
Euclidean metric defined on $D\left(  \Sigma\right)  $ and that in coordinates
(\ref{hds2}) the Euclidean metric has the canonical form, i.e., $diag\left(
+1,+1,+1,+1,+1\right)  $. Then, for example, $\left\|  g^{\ast}-\widetilde
{g},V\right\|  _{m=0}=\left[  \int_{V}\left|  g^{\ast}-\widetilde{g}\right|
^{2}d^{4}xd\psi\right]  ^{\frac{1}{2}}$ where
\[
\left|  g^{\ast}-\widetilde{g}\right|  =\left[
{\textstyle\sum\limits_{i,j=1}^{4}}
\left(  g_{ij}^{\ast}-\widetilde{g}_{ij}\right)  ^{2}+2%
{\textstyle\sum\limits_{i=1}^{4}}
\left(  g_{i5}^{\ast}\right)  ^{2}+\left(  g_{55}^{\ast}-1\right)
^{2}\right]  ^{\frac{1}{2}}%
\]

As we have mentioned, in the given coordinates, the embedding map is $\left(
x,\psi=0\right)  $. Thus the induced metric in $N$ by the new solution
$g^{\ast}$ is given by $g_{ij}^{\prime}\left(  x\right)  =g_{ij}^{\ast}\left(
x,\psi=0\right)  $. Let us show that if $\varepsilon$ is sufficiently small
the induced metric $g^{\prime}$ in $N$ is Lorentzian.

Metrics which are $C^{\infty}$ in the whole manifold belong to local Sobolev
spaces $W^{m}$ for any $m$. Thus they obey some important inequalities which
hold on Sobolev spaces. For example, according to lemma 7.4.1 in ref.
\cite{HE} (page 235), we have that for $m\geq3$, $\left|  g^{\ast}%
-\widetilde{g}\right|  \leq P\left\|  g^{\ast}-\widetilde{g},V\right\|  _{m}$
on $V$, where $P$ is a positive constant (depending on $V$). Thus if $\left\|
g^{\ast}-\widetilde{g},V\right\|  _{m}<\varepsilon$, it follows that all
components satisfy the inequality $\left|  g_{\mu\nu}^{\ast}-\widetilde
{g}_{\mu\nu}\right|  <P\varepsilon$ on $V$.

The gradient $\nabla^{\ast}\psi$ ($\nabla^{\ast}$ is the covariant derivative
compatible with $g^{\ast}$) determines normal direction of the hypersurface
$\psi=0$ in respect to the new solution $g^{\ast}$. The induced metric will be
Lorentzian if $\nabla^{\ast}\psi$ is spacelike with respect to the metric
$g^{\ast}$. In coordinates (\ref{hds2}), we have
\[
g^{\ast}\left(  \nabla^{\ast}\psi,\nabla^{\ast}\psi\right)  =\left(  g^{\ast
}\right)  ^{55}%
\]

Now, this metric component can be written as%
\[
\left(  g^{\ast}\right)  ^{55}=\frac{\det g_{ij}^{\ast}}{\det g_{\mu\nu}%
^{\ast}}.
\]
Since $\det g_{\mu\nu}^{\ast}<0$ on $V$, $\left(  g^{\ast}\right)  ^{55}$ will
be positive if $\det\left(  g_{ij}^{\ast}\right)  <0$. We know that the
determinant is a continuous function with respect to the matrix components in
the usual norm. So there is some $\xi>0$ which depends only on $\widetilde{g}$
and $V$ (an estimate of $\xi$ is given in appendix II) such that for
$\varepsilon<\xi$ we have
\[
\left|  \det g_{ij}^{\ast}-\det\widetilde{g}_{ij}\right|  <\inf_{V}\left|
\det\widetilde{g}_{ij}\right|
\]
This means that $\det g_{ij}^{\ast}<0$ for all points in $V$, since
$\det\left(  \widetilde{g}_{ij}\right)  $ is negative in $V$. This shows that
$g^{\prime}$ is Lorentzian for $\varepsilon<\xi$.

Now let us compare the induced metric $g^{\prime}$ with the original spacetime
metric $g$. It is known that for a field in a Sobolev space the norm of the
restriction of that field to a hypersurface is related to its norm in the
manifold (see \cite{HE}, lemma 7.4.3, page 235). According to the mentioned
lemma 7.4.3, there is a positive constant $Q$ (depending on $N$ and $V$ ) such
that $\left\Vert g^{\ast}-\widetilde{g},N\right\Vert _{m}\leq Q\left\Vert
g^{\ast}-\widetilde{g},V\right\Vert _{m+1}$, where $\left\Vert g^{\ast
}-\widetilde{g},N\right\Vert _{m}$ is defined on $N$ with the induced
Euclidean metric. Thus, for example,%
\[
\left\Vert g^{\ast}-\widetilde{g},N\right\Vert _{0}=\left(  \int_{N}\left\vert
g^{\ast}-\widetilde{g}\right\vert ^{2}d^{4}x\right)  ^{\frac{1}{2}}%
\]
On the other hand $\left\Vert g^{\ast}-\widetilde{g},N\right\Vert _{m}^{\sim
}\leq\left\Vert g^{\ast}-\widetilde{g},N\right\Vert _{m}$ (by $\sim$ we mean
that derivatives are taken only in directions tangent to $N$). Now consider
$g^{\prime},$ the induced metric on $\ N$. In this special coordinates it is
easy to see $\left\Vert g^{\prime}-g,N\right\Vert _{m}^{\sim}$ $\leq\left\Vert
g^{\ast}-\widetilde{g},N\right\Vert _{m}^{\sim}$, since the induced metric has
less components than the higher dimensional metric. Therefore $\left\Vert
g^{\prime}-g,N\right\Vert _{m}^{\sim}\leq Q\left\Vert g^{\ast}-\widetilde
{g},V\right\Vert _{m+1}$.

From the results obtained above we can now show that if the initial data are
sufficiently close to $\left(  h,K\right)  $ on a neighbourhood of the causal
past of $N$ in $\Sigma$ then the induced metric by the new solution is
Lorentzian and close to $g$ in $N$.

More precisely, \textit{let }$N$\textit{ be an open set in }$D^{+}\left(
\Sigma\right)  \cap M$\textit{ with compact closure in }$M$\textit{ and }%
$\eta>0$\textit{, there exist some neighborhood }$\mathcal{U}$\textit{ of
}$J^{-}\left(  \overline{N}\right)  \cap\Sigma$\textit{ of compact closure in
}$\Sigma$\textit{ and some }$\delta>0$\textit{ such that }$C^{\infty}$\textit{
initial data }$\left(  h^{\prime},K^{\prime}\right)  $\textit{ close to
}$\left(  h,K\right)  $\textit{ in }$\mathcal{U}$\textit{, i.e., }$\left\Vert
h^{\prime}-h,\mathcal{U}\right\Vert _{m}^{\sim}<\delta$\textit{ and
}$\left\Vert K^{\prime}-K,\mathcal{U}\right\Vert _{m-1}^{\sim}<\delta
$\textit{, give rise to a metric }$g^{\ast}$\textit{ which induces a
Lorentzian metric in }$N$\textit{ which is near to }$g$\textit{, i.e.,
}$\left\Vert g^{\prime}-g,N\right\Vert _{m-1}^{\sim}<\eta$\textit{.}

In order to see this it suffices to take $m>4$ (\cite{CB}) and to make
appropriate choices for $\varepsilon$ and $V$ in the Cauchy stability theorem.
Indeed, let $V$ be any neighbourhood of $N$ in $D^{+}\left(  \Sigma\right)  $
with compact closure. Since $\overline{N}\subset\overline{V}$ then
$J^{-}\left(  \overline{N}\right)  \subset J^{-}\left(  \overline{V}\right)
$. Thus, taking $\mathcal{U}$ to be some neighbourhood of $J^{-}\left(
\overline{V}\right)  \cap\Sigma$ of compact closure in $\Sigma$, the Cauchy
stability theorem ensures that there exists $\delta$ such that the new metric
generated $g^{\ast}$ satisfies $\left\|  g^{\ast}-\widetilde{g},V\right\|
_{m}<\varepsilon$ for any $\varepsilon>0$. Now if we take $\varepsilon
<\min(\xi$,$\frac{\eta}{Q})$, we guarantee that the induced metric $g^{\prime
}$ is Lorentzian in $N$ and that $\left\|  g^{\prime}-g,N\right\|
_{m-1}^{\sim}<\eta$.

\section{Final Remarks}

From this analysis we can conclude that for each ambient space whose existence
is guaranteed by CM theorem there corresponds an initial data set\emph{
}$\left(  h,K,\Sigma\right)  $\emph{ }in respect to which it possess, in a
certain domain, the desirable physical properties of causality and stability.
This cannot be ensured by CM theorem itself, but by an indirect reasoning as
discussed above. As a direct consequence of this result we have found that for
any analytic spacetime there exist initial data in whose Cauchy development
for the vacuum Einstein equations it can be locally embedded. The embedding is
Cauchy stable and obeys the domain of dependence property with respect to
$\left(  h,K,\Sigma\right)  $.

The extension of the above result to the case where a cosmological constant is
included in the field equations can easily be done. On the other hand the same
analysis cannot be applied to the case of a brane in a straightforward way.
The problem is that in this case we cannot guarantee the existence of a smooth
spacelike hypersurface in an open set of $\widetilde{M},$ at least by the
method employed here, because, as it is known, the brane must be embedded in a
space whose metric has a discontinuity in the derivative along the normal
direction with respect to the brane. Thus this metric is not analytic in any
open set containing the point $p$ and for this reason we cannot use the
Cauchy-Kowalewski theorem to guarantee the existence of a function whose
gradient is everywhere timelike as we have done in the appendix I.
Nevertheless, this question deserves further investigation.

The analyticity of the initial data in $\Sigma$ is a restriction imposed by CM
theorem. However, they are not to be considered an unphysical condition.
Indeed, it must be realized that a great part of the physical solutions, even
in the relativistic regime, are analytic in a certain domain. The crucial
point here is the domain of convergence. What seems to be an unrealistic is a
field which is analytic in the whole manifold. Thus we can say that if we
could handle the initial data in the spacelike four-dimensional hypersurface,
it would be physically feasible to prepare initial data which are analytic in
the interior of a compact domain $S\subset\Sigma$ containing the point $p$ in
order to generate the desired embedding in the interior of $D(S)$.

Moreover we have seen that $C^{\infty}$ initial data sufficiently close to an
analytic initial data set give rise to embeddings of $C^{\infty}$spacetimes
which are near to the original analytic spacetime.\emph{ }This may suggest the
possibility of extending the CM theorem to a less restrictive differential
class of embedddings. We are currently investigating this possibility.

\section{Appendix I}

Consider the five-dimensional vacuum space$\left(  \widetilde{M},\widetilde
{g}\right)  $ into which the spacetime $\left(  M,g\right)  $ is locally
embedded around the point $p\in M$. Let us take the following equation for the
unknown function $\phi:$%
\begin{equation}
\widetilde{g}^{\alpha\beta}\partial_{\alpha}\phi\partial_{\beta}\phi=-1
\end{equation}
In Gaussian coordinates this equation can be written in the following form
\begin{equation}
\frac{\partial\phi}{\partial\psi}=\pm\sqrt{-1+\widetilde{g}^{ij}\partial
_{i}\phi\partial_{j}\phi}%
\end{equation}
Let us consider the equation with the positive sign. According to the
Cauchy-Kowalewski theorem, there is an open set $\widetilde{O}$ in
$\widetilde{M}$ containing the point $p$ where the equation has a solution
provided the initial condition
\begin{equation}
\left.  \phi\right\vert _{\psi=0}=f\left(  x\right)
\end{equation}
ensures that the right-hand side of that equation be an analytic function of
its arguments $\left(  \partial_{i}\phi\right)  $ at the point $p.$ This can
be achieved by choosing $f\left(  x\right)  $ in such a way that the following
inequality be satisfied at the point $p:$%
\begin{equation}
\left.  \left(  g^{ij}\partial_{i}f\partial_{j}f\right)  \right\vert _{p}>1
\end{equation}
This can always be done. For example, take $f\left(  x\right)  =\lambda
V_{i}x^{i}$, where $\lambda>1$ and $V_{i}$ is the component of a unit
spacelike vector with respect to the spacetime metric at the point $p$.

Therefore, the solution $\phi$ is a function whose gradient is everywhere
timelike in $\left(  \widetilde{O},\widetilde{g}\right)  $, and this means
that stable causality condition holds on $\left(  \widetilde{O},\widetilde
{g}\right)  $. Now let us assume without loss of generality that $\phi\left(
p\right)  =0.$ Considering that the gradient of $\phi$ does not vanish in
$\widetilde{O}$, we know that the inverse image $\phi^{-1}\left(  0\right)  $
is a hypersurface $\Sigma$ of $\widetilde{O}$. Since the gradient of $\phi$
which is orthogonal to $\Sigma$ is everywhere timelike, then we can conclude
that $\Sigma$ is a spacelike hypersurface. Moreover, $\Sigma$ is achronal.
Indeed, every future directed timelike curve which leaves $\Sigma$ cannot
return to $\Sigma$ since $\phi$ does not change sign along these curves. It
happens that an achronal spacelike hypersurface is acausal \cite{oneill}.
Therefore $\Sigma$ is an acausal spacelike hypersurface in $\left(
\widetilde{O},\widetilde{g}\right)  $.

\section{\bigskip Appendix II}

In this section we want to determine an estimate of how much near to
$\widetilde{g}$ the new metric $g^{\ast}$ must be in order to induce a
Lorentzian metric in $N$. As described above, this is achieved if the
condition%
\[
\left|  \det g_{ij}^{\ast}-\det\widetilde{g}_{ij}\right|  <\inf_{V}\left(
\det\widetilde{g}_{ij}\right)
\]
holds on $V$. As we have seen, for $m\geq3$, $\left\|  g^{\ast}-\widetilde
{g},V\right\|  _{m}<\varepsilon$ implies that $\left|  g_{\mu\nu}^{\ast
}-\widetilde{g}_{\mu\nu}\right|  <P\varepsilon$ on $V$. Then, there exist a
continuous field $\gamma_{\mu\nu}$, with $\left|  \gamma_{\mu\nu}\right|  <1$
on $V,$ such that $g_{\mu\nu}^{\ast}=\widetilde{g}_{\mu\nu}+\theta\gamma
_{\mu\nu}$, where $\theta=P\varepsilon$. We want to evaluate $\det
g_{ij}^{\ast}$. First let us define $\omega^{ijkm}$ as a totally
anti-symmetric four index `tensor' with $\omega^{1234}=1$. Thus, we can write%
\[
\det g_{ij}^{\ast}=\omega^{ijkm}\left(  \widetilde{g}_{1i}+\theta\gamma
_{1i}\right)  \left(  \widetilde{g}_{2j}+\theta\gamma_{2j}\right)  \left(
\widetilde{g}_{3k}+\theta\gamma_{3k}\right)  \left(  \widetilde{g}_{4m}%
+\theta\gamma_{4m}\right)
\]
This expression can be rewritten in the following form%
\[
\det g_{ij}^{\ast}=\det\widetilde{g}_{ij}+\theta I_{1}+\theta^{2}I_{2}%
+\theta^{3}I_{3}+\theta^{4}\det\gamma_{ij}%
\]
where%
\begin{align*}
I_{1}  &  =\omega^{ijkm}\left(  \widetilde{g}_{1i}\widetilde{g}_{2j}%
\widetilde{g}_{3k}\gamma_{4m}+\widetilde{g}_{1i}\widetilde{g}_{2j}\gamma
_{3k}\widetilde{g}_{4m}+\widetilde{g}_{1i}\gamma_{2j}\widetilde{g}%
_{3k}\widetilde{g}_{4m}+\gamma_{1i}\widetilde{g}_{2j}\widetilde{g}%
_{3k}\widetilde{g}_{4m}\right) \\
I_{2}  &  =\omega^{ijkm}\left(  \widetilde{g}_{1i}\widetilde{g}_{2j}%
\gamma_{3k}\gamma_{4m}+\widetilde{g}_{1i}\gamma_{2j}\widetilde{g}_{3k}%
\gamma_{4m}+\gamma_{1i}\widetilde{g}_{2j}\widetilde{g}_{3k}\gamma
_{4m}+\widetilde{g}_{1i}\gamma_{2j}\gamma_{3k}\widetilde{g}_{4m}\right. \\
&  \hspace{0.75in}\left.  +\gamma_{1i}\widetilde{g}_{2j}\gamma_{3k}%
\widetilde{g}_{4m}+\gamma_{1i}\gamma_{2j}\widetilde{g}_{3k}\widetilde{g}%
_{4m}\right) \\
I_{3}  &  =\omega^{ijkm}\left(  \widetilde{g}_{1i}\gamma_{2j}\gamma_{3k}%
\gamma_{4m}+\gamma_{1i}\widetilde{g}_{2j}\gamma_{3k}\gamma_{4m}+\gamma
_{1i}\gamma_{2j}\widetilde{g}_{3k}\gamma_{4m}+\gamma_{1i}\gamma_{2j}%
\gamma_{3k}\widetilde{g}_{4m}\right)
\end{align*}

\bigskip Therefore, we have%

\[
\left\vert \det g_{ij}^{\ast}-\det\widetilde{g}_{ij}\right\vert \leq
\theta\left\vert I_{1}\right\vert +\theta^{2}\left\vert I_{2}\right\vert
+\theta^{3}\left\vert I_{3}\right\vert +\theta^{4}\left\vert \det\gamma
_{ij}\right\vert
\]

In order to make estimates of $I_{1},I_{2},I_{3}$ it is necessary to eliminate
$\gamma_{ij}$. To this end, let us introduce $\chi_{ij},$ a $4x4$ matrix with
all elements equal to unity. Using this matrix, we get the following estimate%
\[
\left\vert \omega^{ijkm}\widetilde{g}_{1i}\widetilde{g}_{2j}\widetilde{g}%
_{3k}\gamma_{4m}\right\vert \leq\left\vert \omega^{ijkm}\right\vert \left\vert
\widetilde{g}_{1i}\widetilde{g}_{2j}\widetilde{g}_{3k}\gamma_{4m}\right\vert
<\left\vert \omega^{ijkm}\right\vert \left\vert \widetilde{g}_{1i}%
\widetilde{g}_{2j}\widetilde{g}_{3k}\chi_{4m}\right\vert
\]
Similar inequalities hold for every term that appear in $I_{1},I_{2},I_{3}$.
Then we have $\left\vert I_{s}\right\vert <\left\vert J_{s}\right\vert $, for
$s=1,2,3$, where $\left\vert J_{s}\right\vert $ is obtained from $I_{s}$ by
substituting each one of its term for the corresponding estimate as above. It
happens that $\left\vert J_{s}\right\vert $ depends only on $\widetilde
{g}_{ij}$. Indeed, it is a function constituted of sum and multiplication of
the components of $\widetilde{g}_{ij}$. Now for the sake of simplicity let us
consider that $\theta<1$, (i.e., $\varepsilon<\frac{1}{P}$), then we have%
\[
\left\vert \det g_{ij}^{\ast}-\det\widetilde{g}_{ij}\right\vert <P\varepsilon
\left(  \left\vert J_{1}\right\vert +\left\vert J_{2}\right\vert +\left\vert
J_{3}\right\vert +4!\right)  ,
\]
recalling that $\left\vert \det\gamma_{ij}\right\vert <4!$. Since
$\widetilde{g}_{ij}$ is continuous in $\widetilde{M}$, then the right hand
side is bounded in any compact set such as $\overline{V}$. The same is valid
for $\det\widetilde{g}_{ij}$. Now define $\xi_{1}$ as the positive quantity%
\[
\xi_{1}=\frac{\underset{V}{\inf}\left\vert \det\widetilde{g}_{ij}\right\vert
}{P\left[  \underset{V}{\sup}\left(  \left\vert J_{1}\right\vert +\left\vert
J_{2}\right\vert +\left\vert J_{3}\right\vert +4!\right)  \right]  }%
\]
The value of $\xi_{1}$ depends only on $\widetilde{g}$ and $V$ and the chart
(possibly one can give a definition independent of coordinates, but this
definition is sufficient for our purpose). Now define $\xi=\min\left\{
\frac{1}{P},\xi_{1}\right\}  .$ Taking $\varepsilon<\xi,$ we get%
\[
\left\vert \det g_{ij}^{\ast}-\det\widetilde{g}_{ij}\right\vert <\underset
{V}{\inf}\left\vert \det\widetilde{g}_{ij}\right\vert
\]
Considering that $\det\widetilde{g}_{ij}<0$ on $\overline{V},$ we conclude
that $\det g_{ij}^{\ast}<0$ on $V$ and hence the induced metric in $N$ is Lorentzian.

\section{Acknowledgements}

The authors thank CNPq for financial support. This work was partially financed
by CNPq/FAPESQ (PRONEX).

\end{document}